\documentclass[12pt,reqno]{amsart}
\usepackage{amsmath,amssymb, amscd}

\usepackage[pdftex]{graphicx}
\usepackage{amscd}
\usepackage{graphics}

\makeatletter
\@addtoreset{equation}{section}
\makeatother
\usepackage{enumerate}

\usepackage{color}

\newcommand{\revs}[1]{{#1}}

\def\CC{{\mathbb C}}

\def\RR{{\mathbb R}}

\def\ee{{\mathrm e}}

\def\ii{{\mathfrak{i}}}

\def\cE{{\mathcal E}}

\def\cI{{\mathcal I}}

\def\tOmega{{\widetilde \Omega}}

\def\SO{\mathrm{SO}}

\def\UU{\mathrm{U}}

\def\vol#1{\textbf{#1}}

\def\dfrac#1#2{{\displaystyle\frac{#1}{#2}}}

\def\vol#1{\textbf{#1}}

\begin{document}

\title{Euler's original derivation of elastica equation}
\date{\today}

\author{Shigeki Matsutani}



\begin{abstract}
Euler derived the differential equations of elastica by the variational method in 1744, but his original derivation has never been properly interpreted or explained in terms of modern mathematics.
We elaborate Euler's original derivation of elastica and show that Euler used Noether's theorem concerning the translational symmetry of elastica, although Noether published her theorem in 1918.
It is also shown that his equation is essentially the static modified KdV equation which is obtained by the isometric and isoenergy conditions, known as the Goldstein-Petrich scheme.
\end{abstract}

\maketitle
\bigskip



\section{Introduction}
{\revs{
The elastica problem is to mathematically represent and classify the shapes of the thin ideal elastic rod, which was proposed by Jacob Bernoulli (1654--1705) \cite{T1}.
It is well-known that 
}
Euler (1707--1783) derived the differential equations of the elastica\revs{, elastic curve,} by the variational method and \revs{completely} classified the moduli of elasticae in 1744 \cite{E2}.
However, his original derivation has never been properly interpreted or explained in terms of modern mathematics.
\revs{
Bolza (1857--1942), who was the expert of the isoperimetric problems of the variational method theory and the Abelian function theory including the elliptic integrals \cite{M3}, had the opportunity to survey Euler's work of the elastica in the review of Gauss's (1777--1855) elastica but he could not analyze Euler's elastica well \cite{B}. 
}
Since Euler's derivation is not \revs{superficially} difficult to follow, anyone can trace it.
However, its mathematical or physical meaning is not clear.
In other words, no one gave any particular comment, without formulae, on what Euler did and how he did his derivation as far as we know.

Even Truesdell gave a only short comment on his derivation in \cite{T1}, {\lq\lq}since $Pdp = dZ - Q dq$, another is easy, yielding $\alpha \sqrt{1+p2}+\beta p + \gamma = Z- Qq.${\rq\rq} This equation is the same as (\ref{eq:1-6}) in this paper, which is essential in Euler's derivation.
Further, D'Antonio, Levien, and Bistafa gave excellent reviews of Euler's derivation respectively, but did not give any particular interpretation on (\ref{eq:1-6}) from a mathematical point of view \cite{DA, Le,Bi}.

As recently, Dittrich explained Euler's elastica from a viewpoint of modern theoretical physics using the quantities appearing in Euler's derivation, his explanation is proper and nice as theory of elastica but does not correspond to Euler's original derivation at all \cite{Di}.

\begin{figure}[htb]
\begin{center}
\includegraphics[width=0.4\hsize]{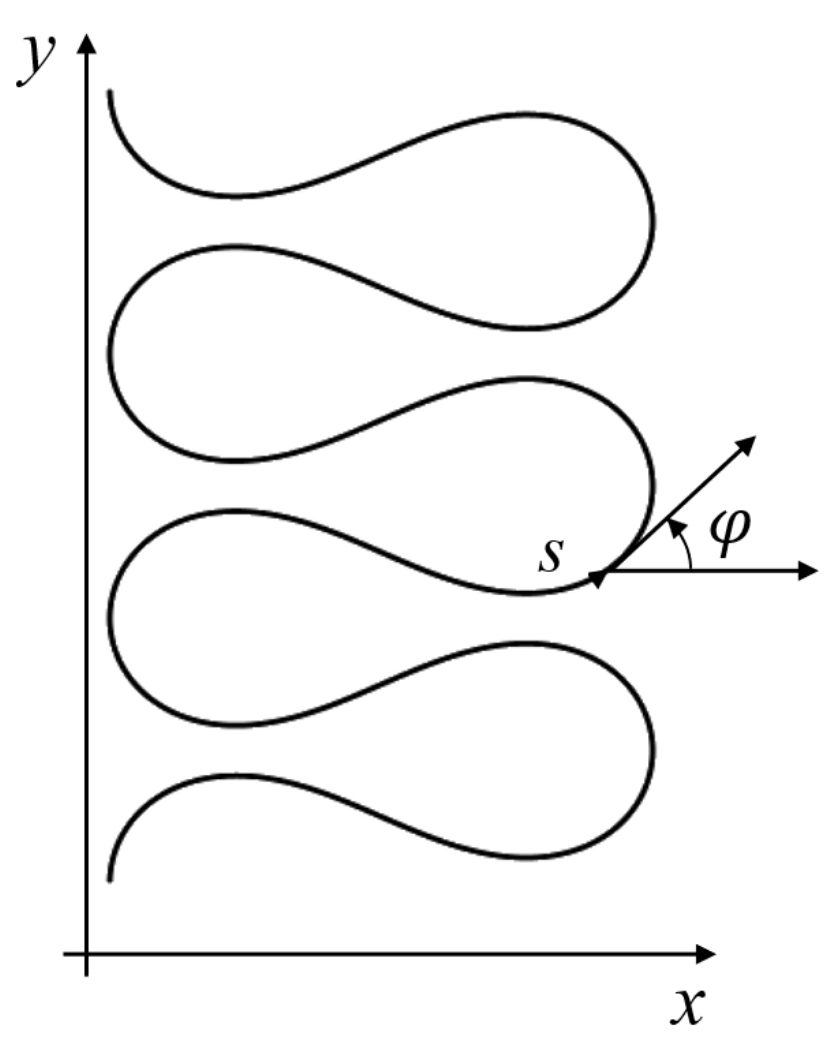}\\
\caption{Shape of elastica: $s$ is the arclength and $\varphi$ is the tangential angle}
\label{fg:1.1}
\end{center}
\end{figure}

In fact, Euler's original formula is different from the well-known equation of elastica \cite{Love}, i.e. the static sine-Gordon equation or the nonlinear pendulum equation, 
\begin{equation}
    \frac{d^2 \varphi}{d s^2} = a \sin \varphi,
\label{eq:SGe}
\end{equation}
where $\varphi$ is the tangential angle and $s$ is the arclength of elastica \revs{illustrated in Figure \ref{fg:1.1}},
which was obtained by Lagrange (1736--1813) and Gauss \cite{La, Ga}\revs{,} (See Appendix \revs{A.1.)}

\revs{
In the study of the integrable system of nonlinear partial differential equations after 1950's, the pendulum equation (\ref{eq:SGe}) was reinterpreted as the static (ordinary differential) version of the sine-Gordon equation \cite{AS},
\begin{equation}
   \frac{\partial^2 \varphi}{\partial t^2}- \frac{\partial^2 \varphi}{\partial s^2} = a \sin \varphi,
\label{eq:TSGe}
\end{equation}
where $t$ is the time parameter.
If $\varphi$ is constant in $t$, (\ref{eq:TSGe}) becomes (\ref{eq:SGe}).
Since Euler's original equation of elastica is treated in a similar way later on, we refer to (\ref{eq:SGe}) as the static sine-Gordon equation in this paper.
}

The derivation of (\ref{eq:SGe}) is very simple; it can be obtained by a straightforward way \revs{by minimizing the energy functional $\displaystyle{\int \kappa^2 ds}$ under the isometric \revs{(non-stretching, isoperimetric)} condition as Gauss obtained (c.f. (\ref{I.6.11}) in Appendix A.1). 
Here $\kappa$ is the curvature $\displaystyle{\kappa = \frac{d \varphi}{d s}}$ of the elastica, and is the inverse of the curvature radius $R(s)$, $\kappa(s) = 1/R(s)$.
It is known that the energy functional was discovered by Daniel Bernoulli (1700--1782) \cite{DA, Le,Bi, T1} and it is called the Euler-Bernoulli energy functional.
Euler referred to the non-stretching condition as the isoperimetric condition, although from the modern mathematical point of view, it should be called the isometric condition.
Due to this discovery, the elastica problem is regarded as the minimization of the energy under the isometric (non-stretching, isoperimetric) condition.
}

Thus in most of \revs{the} essays of \revs{the} elastica \cite{T2, Love}, (\ref{eq:SGe}) appears as a governing equation of elastica. 
Dittrich also showed the relation between (\ref{eq:SGe}) and the quantities appearing in Euler's derivation \cite{Di}.
However, since \revs{(\ref{eq:SGe})} did not appear in Euler's original derivation, Dittrich's investigation is not correct from the point of view of the historical interpretation of Euler's derivation.

\bigskip

\revs{
In the study of the integrable nonlinear partial differential equations \cite{AS}, the modified Kortweg-de Veries equation,
\begin{equation}
   \frac{\partial \varphi}{\partial t}
+ \frac{1}{2}\left[ \frac{\partial \varphi}{\partial s}\right]^3
+ \frac{\partial^3 \varphi}{\partial s^3} =0,
\label{eq:MKdVeq0}
\end{equation}
appeared as integrable models of several physical phenomena, including a plane curve \cite{KIW, Ishimori, Ishimori2}: Euler's elastica was rediscovered in the framework around 1980 \cite{KIW, Ishimori, Ishimori2}.
}

As mentioned in \cite{BG, LS, M1}, the elastica also obeys \revs{the equation,}
 \begin{equation}
 a  \frac{d \varphi}{d s}+ \frac{1}{2}\left[ \frac{d \varphi}{d s}\right]^3
+ \frac{d^3 \varphi}{d s^3} =0,
\label{eq:SMKdVeq0}
\end{equation}
\revs{
where $a$ is a certain real constant number, which is obviously different from the static sine Gordon equation (\ref{eq:SGe}).
Since (\ref{eq:SMKdVeq0}) is obtained by setting $t=s/a$ in (\ref{eq:MKdVeq0}), we refer to (\ref{eq:SMKdVeq0}) as the static modified KdV (SMKdV) equation, although \cite{BG} and \cite{LS} did not call it the SMKdV equation.}
\revs{Since} the derivation of the SMKdV equation is, a little bit, complicated due to the isometric condition (see Appendix \revs{A.2), (\ref{eq:SMKdVeq0})} appeared after the middle of the 20th century.
Euler's original equation should be regarded as the SMKdV equation (\ref{eq:SMKdVeq0}) instead of (\ref{eq:SGe}) \revs{by setting $\displaystyle{\kappa=\frac{d\varphi}{ds}}$ proportional to $x$ as in (\ref{eq:1-16}).
(In his derivation, Euler implicitly showed that $\kappa$ is proportional to $x$.)}
\revs{
Like the derivation of (\ref{eq:SMKdVeq0}),} his derivation is also complicated \revs{and difficult to interpret without knowledge of modern mathematics; more precisely, Euler's is much more difficult due to translational symmetry, as we will mention below.}

Hence, Euler's derivation should be interpreted from a modern viewpoint\revs{.} \revs{The} SMKdV equation is very natural from a modern mathematical viewpoint, and his derivation is associated with the recent discovery based on the Goldstein-Petrich scheme \cite{GP, MP} (see Appendix \revs{A.3):
Euler's result is so profound and natural, although none of his successors understood it.
}

\bigskip

In this paper, we elucidate Euler's original derivation of elastica, and show that Euler used Noether's theorem related to the translational symmetry of the elastica, and derived the SMKdV equation \revs{(\ref{eq:SMKdVeq0})} rather than the static sine-Gordon equation (\ref{eq:SGe}), although Noether published her theorem in 1918 \cite{N}.

\revs{
Here we give a short explanation on the translational symmetry in the elastica.
Let us consider the isometric immersion $z$ of a loop $S^1$ in a real plane $\RR^2 = \CC$, $z : S^1 \to \CC$; $z(s)=x(s)+\ii y(s)$, $z(s+2\pi)=z(s)$, where $s$ is the arclength of the loop. (We note that {\lq\lq}immersion{\rq\rq} is a terminology in mathematics and thus, roughly speaking, it means that we consider a closed curve $\{z(s)\}$ in $\CC$, which we allow several simple crossings).
Then we encounter a freedom in how we set the origin of the coordinate system (or the arclength). 
Mathematically, it corresponds to a trivial group action on $\{z(s)\}$ of the unitary group $\UU(1):=\{g_s:=\ee^{\ii s}\ | \ s\in \RR \}$: for $g_{s_0}$ and $g_{s_1}\in \UU(1)$, $g_{s_0} g_{s_1}=g_{s_0+s_1}\in \UU(1)$, and the action $\Phi_{g_{s'}}$ on the loop $\{z\}$ by  $g_{s'}\in \UU(1)$ is defined by $\Phi_{g_{s'}} z(s)=z(s+s')$.
Since a set with a group action is said to have a symmetry in mathematics, we refer to the group action or the freedom as the translational symmetry.
Then the origin $z(0)$ of $s$ is replaced by $\Phi_{g_{s}}z(0) = z(s)$ by the $\UU(1)$ action due to the translational symmetry. 
In other words, {\it{any point on the curve $\{z\}$ can be the origin of the coordinate $s$.}}
The translational symmetry is the most essential symmetry of the immersed closed loops.
(It is regarded as a gauge symmetry in theoretical physics.)
}

\revs{
The symmetry also exists for the isometric immersion of the curves with the infinite length, and of the curve with the periodic properties $\displaystyle{\frac{d z}{ds}(s+L) = \frac{d z}{ds}(s)}$, which Euler implicitly assumed: (the elastica has the periodic properties as in (\ref{eq:wp}).)
Therefore, in the problem of elastica, the translational symmetry should be noticed although there have been no arguments on the symmetry except by Euler, before 1900.
}

It is noted that when we consider the generalization of elastica, or the excited states of the Euler-Bernoulli energy functional, the SMKdV equation  (\ref{eq:SMKdVeq0}) is more natural than the static sine-Gordon equation (\ref{eq:SGe}), due to the symmetry \cite{MP}.

The modern derivation of the SMKdV equation as the governing equation of elastica is based on the non-stretching condition with \revs{the} translational symmetry that Euler focused on\revs{;}
Goldstein and Petrich found a simple but deep representation of the non-stretching condition \cite{GP}, and using it, we can find the SMKdV equation satisfying the translational symmetry \revs{by considering the static version (\ref{eq:SMKdVeq0}) of the MKdV equation (\ref{eq:MKdVeq0}) as we show in Appendix A.3.}
This view existed in Euler's original derivation.
\revs{
Euler considered the variational problem under the isoperimetric (non-stretching, isometric) condition with the translational symmetry.
}

Further, recently we show that the symmetry is also related to the shapes of supercoiled DNAs \cite{M24}.
Thus, in this stage, it is very important to elucidate what Euler did in his derivation of the elastica \revs{equation}.

Then we must say that Euler revealed the essence of elastica.

\bigskip

The contents are as follows.
Section 2 is devoted to the explanation of Euler's deviation of his elastica equation.
The mathematical significance of his derivation is discussed in Section 3.
We include Appendix \revs{A.1,} where we show Lagrange's equation of elastica and Gauss' derivation.
We also show that the elastica obeys the SMKdV equation \revs{in Appendices A.2 and A.3}.

\section{Euler's derivation of the elastica equation in 1744}

Following Daniel Bernoulli's suggestion, in the Appendix to \cite{E2}, Euler derived the elastica equation, which he had already obtained in 1732 \cite{E1}, by means of the variational method \revs{in 1744} \cite{DA,Bi}.

In this section, we show Euler's derivation of the elastica equation in \cite[pp.247-250]{E2}\revs{\cite[pp.79-82]{OEB}} and explain it from a modern mathematical point of view.

Under \revs{the} boundary conditions with fixed ends \revs{$A$} and \revs{$B$}, Euler considered the problem of determining the curve, i.e., an elastica, that is the minimum of the Euler-Bernoulli functional $\displaystyle{\int \frac{1}{R^2} \frac{ds}{dx} dx=\int Z dx}$ \revs{so that it satisfies the isometric (isoperimetric) condition.}
Here $R$ is the curvature radius of the elastica\revs{, the inverse of the curvature $\displaystyle{\frac{d \varphi}{d s}=\frac{1}{R}}$,} and $s$ is the arclength of the elastica, which are regarded as multiple-valued functions on $x$ \revs{ as illustrated in Figure \ref{fg:1.1}}.

From a modern point of view, he considered a curve $\{(x(s),y(s))\ | \ s\in [0, L] \}$, $(L>0)$ 
and \revs{
implicitly assumed that $\displaystyle{\left(\frac{d x(s)}{ds}, \frac{d y(s)}{ds}\right)=\left(\frac{d x(s+L)}{ds},\right.}$
$\displaystyle{\left. \frac{d y(s+L)}{ds}\right)}$ as in Introduction, although he did not mentioned at all.
He} also regarded $y(x)$ as a multiple-valued function on $x$,
\revs{
and as the results, he derived $x(s+L)=x(s)$ as in Figure \ref{fg:1.2}.
}

As in \cite[p.248]{E2}\revs{\cite[p.79]{OEB}}, Euler introduced the parameters $p:=\displaystyle{\frac{dy}{dx}}$ and $q:=\displaystyle{\frac{dp}{dx}}$.
Since $ds = \sqrt{1+ p^2} dx$ and the curvature $\kappa=\displaystyle{\frac{1}{R}}$ is $\displaystyle{\frac{q}{(1+p^2)^{3/2}}}$,
$\displaystyle{Z=\frac{q^2}{(1+p^2)^{5/2}}}$. 
\revs{
He considered the minimum of $\displaystyle{\int Z dx}$ under the isoperimetric condition that $\displaystyle{\int ds=\int \sqrt{1+p^2} dx}$ is constant and under the boundary condition, i.e., the above periodic condition at the fixed points $y(x_0)=y_0$ and $y(x_1)=y_1$ if we set $A=(x_0, y_0)$ and $B=(x_1, y_1)$.
}

He \revs{considered the variation of $Z$} as
\begin{equation}
dZ = M dx + N dy + P dp + Q dq,
\label{eq:1-1}
\end{equation}
where $M=N=0$, $\displaystyle{P = \frac{-5 pq^2}{(1+ p^2)^{7/2}}}$ and 
$\displaystyle{Q = \frac{2q}{(1+ p^2)^{5/2}}}$.

From a modern viewpoint, \revs{if we introduce $U[p]:=\displaystyle{\int Z\left[p,\frac{d p}{dx}\right] dx}$,} he considered the functional (G\^{a}teaux) derivative \revs{\cite{Gat}\cite[p45]{Tro},
$$
U[p; v]:=\lim_{\varepsilon \to 0} 
\frac{U[p+\varepsilon v] - 
U[p]}{\varepsilon}
$$
for a certain function $v$ on $[0, L]$, 
 and thus \cite[p.56]{Tro},
$$
P = \frac{\partial Z[p,q]}{\partial p}, \quad
Q = \frac{\partial Z[p,q]}{\partial q}.
$$
}
In this paper, we sometimes use $\dot{p}$ instead of $q$ by following the modern convention, though Euler did not use it.

Euler went on to investigate (\ref{eq:1-1}) for including the isometric (non-stretching\revs{, isoperimetric}) condition.
\revs{He compared $\displaystyle{\frac{d}{dx}\left[\frac{\partial }{\partial p}\sqrt{1+ p^2}\right]}$ with the differential of the variation of $\displaystyle{\int Z ds}$ with respect to $x$, i.e., $\displaystyle{
-\frac{d P}{dx}+\frac{d^2 Q}{d x^2}}$, so that the isometric condition is preserved.
With a constant factor $\alpha$, he wrote \cite[l.-11, p.248]{E2}\revs{\cite[l.5, p.80]{OEB}},
\begin{equation}
\alpha \frac{d}{dx}\frac{p }{\sqrt{1+ p^2}}
= \frac{d P}{dx}-\frac{d^2 Q}{d x^2}.
\end{equation}
It should be noted that he used the so-called Lagrange multiplier $\alpha$ to handle the isometric condition.
}
Then he introduced the integral constant $\beta$, and had the equation \cite[l.-9, p.248]{E2}\revs{\cite[l.7, p.80]{OEB}},
\begin{equation}
\frac{\alpha p}{\sqrt{1+pp}}+\beta = P - \frac{d}{dx}Q.
\label{eq:r-7}
\end{equation}
\revs{
As in (\ref{eq:1-4}), $\beta$ in (\ref{eq:r-7}) is regarded as the Lagrange multiplier of the boundary condition $y(x_0)=y_0$ and $y(x_1)=y_1$.
Although he did not write (\ref{eq:r-7}) down explicitly, (\ref{eq:r-7}) is equal to
}
\begin{equation}
\frac{-5 p q^2}{(1+p^2)^{7/2}}- \frac{\alpha p}{\sqrt{1+p^2}}
-\frac{d }{d x}\frac{2 q}{(1+p^2)^{5/2}}-\beta=0.
\label{eq:1-2}
\end{equation}
He also regard \revs{(\ref{eq:r-7})} as \cite[l.-8,p.248]{E2}\revs{\cite[l.9, p.80]{OEB}}
$$
\frac{\alpha p dp}{\sqrt{1+pp}}+\beta dp  = P dp- q dQ,
$$
or
\begin{equation}
Pdp- \frac{\alpha pdp }{\sqrt{1+p^2}}-
q d Q-\beta dp=0
\label{eq:1-2a}
\end{equation}
by multiplying \revs{by} $dp$.
Here \revs{Euler used} that $q dx = dp$ and $d Q=\displaystyle{\frac{d}{dx} Qdx}$.

\revs{Since} he considered the fixed boundary conditions at both ends $(x_0, y_0)$ and $(x_1, y_1)$, \revs{and the isoperimetric condition, from a modern viewpoint, he dealt with the functional with} the Lagrangian multipliers $\alpha$ and $\beta$,
\begin{equation}
E[y(x)]:= \int_{x_0}^{x_1} \frac{1}{R^2} ds 
- \alpha \left[L-\int_{x_0}^{x_1} ds\right]
 -\beta \left[\revs{(y_1-y_0)}-\int_{x_0}^{x_1} dy\right],
\label{eq:1-3}
\end{equation}
which is equal to
\begin{equation}
E[y(x)]=
 \int \frac{\dot p^2}{(1+p^2)^{5/2}} dx
-\alpha \int \sqrt{1+ p^2} dx
-\beta\int p dx+
\mbox{constant}.
\label{eq:1-4}
\end{equation}
\revs{
(More precisely, Euler assumed $\displaystyle{\frac{d x}{ds}(s+L)=\frac{d x}{ds}(s)}$, $\displaystyle{\frac{d y}{ds}(s+L)=\frac{d y}{ds}(s)}$:
as functions of $s$, $p(s+L)=p(s)$ and $q(s+L)=q(s)$.)
}
Let us write this integral function as $\cE(p, \dot p)$, i.e., 
$E=\displaystyle{\int \cE(p, \dot p) dx}$\revs{, i.e.,
\begin{equation}
\cE(p,\dot p)=\frac{\dot p^2}{(1+p^2)^{5/2}} 
-\alpha \sqrt{1+ p^2} -\beta p .
\label{eq:1-4b}
\end{equation}
}

By following the Euler-Lagrange equation \cite{Go},
\begin{equation}
\frac{\partial \cE}{\partial p}-\frac{d }{d x} \frac{\partial \cE}{\partial \dot p}=0,
\label{eq:1-4a}
\end{equation}
we recover the equation (\ref{eq:1-2}) which Euler derived, i.e.,
\begin{equation}
\frac{-5 p\dot p^2}{(1+p^2)^{7/2}}- \frac{ \alpha p}{\sqrt{1+p^2}}
-\frac{d }{d x}\frac{2 \dot p}{(1+p^2)^{5/2}}-\beta=0.
\label{eq:1-2p}
\end{equation}
However, Euler did not solve (\ref{eq:1-2}) or (\ref{eq:1-2p}) directly.

Euler focused on the fact that $M=N=0$.
In other words, he \revs{considered} the symmetry in the elastica, which shows the translational invariant.
Euler stated that $dZ = P dq + Q dq$ or $Pdp = dZ - Qdq$ in his notation, whereas from (\ref{eq:1-2a})\revs{,} $Pdp$ is equal to
\begin{equation}
 \frac{2\alpha pdp }{\sqrt{1+p^2}}+
q d Q+\beta dp.
\label{eq:1-2ap}
\end{equation}
Then he reached the first result \cite[l.-5, p.248]{E2}\revs{\cite[l.-10, p.80]{OEB}},
\begin{equation}
\frac{\alpha pdp }{\sqrt{1+pp}}
+\beta dp = dZ - Qdq - q d Q.
\label{eq:1-5}
\end{equation}
He integrated (\ref{eq:1-5}) and obtained \cite[l.-4, p.248]{E2} \revs{\cite[l.-8, p.80]{OEB}},
$$
\alpha \sqrt{1+pp}
+\beta p + \gamma = Z - Qq,
$$
and \revs{\cite[l.-2, p.248]{E2}\cite[l.-5, p.80]{OEB}},
\begin{equation}
\alpha \sqrt{1+pp}
+\beta p + \gamma = -\frac{q q}{(1+pp)^{5:2}}.
\label{eq:1-6}
\end{equation}

\bigskip

However (\ref{eq:1-5}) and (\ref{eq:1-6}) seemed to be very difficult for his successors, although the transformation of the equations is \revs{superficially} simple.
Nobody could understand what they meant.
Lagrange essentially found another equation (\ref{eq:SGe}) as the equation for the elastica \cite{La} and Gauss also followed it \cite{Ga} (see Appendix \revs{A.1}).

Euler considered the translational symmetry with respect to the $x$ direction, which is a symmetry determined by the indefiniteness of the origin of the arc length \revs{$s$} of the curve in the modern sense \revs{though the implicit function theorem.}
It corresponds to $M=N=0$.
Then he derived a conserved quantity, known today as the Noether current \cite[13.7]{Go}\cite{N}, which is determined by the so-called Noether's theorem.
It is equivalent to the momentum map in differential geometry \cite{AM}.

Under (\ref{eq:1-3}) and (\ref{eq:1-2p}), the Noether\revs{'s theorem} in the $x$ direction \revs{shows} the formula \revs{(\ref{eq:1-6})} \cite{Go,AM}.
\revs{
In other words we apply the formula from Noether's Theorem (e.g., (13.28) or (13.158) in \cite{Go}) to this system.
However since Noether's Theorem for $x$-direction is very simple, we derive the  current following the computations in the theorem, e.g., (13.28) in \cite{Go}, as follows.
}

\revs{
By noting the fact that the dependence of $\cE$ with respect to $t$ is only through $p$ and $\dot p$, we consider the differential of $\cE$ with respect to $x$,
\begin{equation}
\frac{d}{dx}\cE = 
\frac{d p}{dx} \frac{\partial \cE}{\partial p}
+\frac{d \dot p}{dx} \frac{\partial \cE}{\partial \dot p}
\end{equation}
By using (\ref{eq:1-4a}), we have
\begin{equation}
\frac{d}{dx}\cE = 
\dot p \frac{d}{d x} \frac{\partial \cE}{\partial \dot p}
+\frac{d \dot p}{dx} \frac{\partial \cE}{\partial \dot p}.
\end{equation}
and thus we obtain
}
\begin{equation}
\frac{d }{d x}\left[\frac{\partial \cE}{\partial \dot p}\cdot \dot p\right]
-\frac{d }{d x}\cE=0.
\label{eq:1-7}
\end{equation}
It means that there is a conserved quantity or the Noether current,
\begin{equation}
\cI:=\left[\frac{\partial \cE}{\partial \dot p}\cdot \dot p\right]-\cE\revs{,
\quad \frac{d}{dx}\cI=0.}
\label{eq:1-8}
\end{equation}
\revs{
Since $\frac{d}{d s} \cI=0$, $\cI$ is constant with respect to $x$, which is referred to a conserved quantity.
}
Writing down (\ref{eq:1-8}) with a constant $\gamma$, \revs{i.e., $\cI+\gamma=0$,} we obtain the conserved current $\cI$ \revs{from (\ref{eq:1-4b})},
\begin{equation}
\frac{\dot p^2}{(1+p^2)^{5/2}}+\alpha\sqrt{1+p^2}+\beta p+\gamma=0,
\label{eq:1-9}
\end{equation}
which is identical to (\ref{eq:1-6}).
\revs{
The derivation of (\ref{eq:1-7}) in \cite[(13.28)]{Go} is basically the same as Euler's above}

In other words, Euler found and used the Noether current (\ref{eq:1-6}) and (\ref{eq:1-9}) even only for the translational symmetry in 1744, while Noether published her paper on \revs{Noether's} theorem for a general group action in 1918 \cite{N}.

\bigskip

Euler made $\alpha$, $\beta$ and $\gamma$ be $-\alpha$, $-\beta$ and $-\gamma$ since they are arbitrary constants.
From (\ref{eq:1-6}) and (\ref{eq:1-9}), he obtained
\begin{equation}
\frac{dp}{dx}=(1+p^2)^{5/4}\sqrt{(\alpha\sqrt{(1+p^2)}+\beta p+\gamma}.
\label{eq:1-10}
\end{equation}
By noting that $dy = p dx$, Euler found \cite[l.3, l.5, p.249]{E2}
 \revs{\cite[l.-1, p.80, l.2, p.81]{OEB}}
\begin{equation}
dx = \frac{dp}{
(1+pp)^{5:4}\sqrt{\alpha\sqrt{(1+pp)}+\beta p+\gamma}},
\label{eq:1-10a}
\end{equation}
\begin{equation}
dy = \frac{pdp}{
(1+pp)^{5:4}\sqrt{\alpha\sqrt{(1+pp)}+\beta p+\gamma}}.
\label{eq:1-11}
\end{equation}

Further, Euler noticed the relation \cite[l.10, p.249]{E2} \revs{\cite[l.8, p.81]{OEB}},
\begin{equation}
d
\frac{2\sqrt{\alpha\sqrt{(1+pp)}+\beta p+\gamma}}
{\sqrt{\sqrt{1+pp}}}
=
\frac{dp (\beta-\gamma p)}
{(1+pp)^{5:4}\sqrt{\alpha\sqrt{(1+pp)}+\beta p+\gamma}}.
\label{eq:1-12}
\end{equation}
By using the relation, he derived \cite[l.11, p.249]{E2} \revs{\cite[l.9, p.81]{OEB}}
\begin{equation}
\frac{2\sqrt{\alpha\sqrt{(1+pp)}+\beta p+\gamma}}
{(1+pp)^{1:4}}=\beta x-\gamma y+\delta. 
\label{eq:1-12a}
\end{equation}

\revs{
Euler investigated the (global) euclidean symmetry of $(x, y)$, which consists of the translational symmetry and the rotational symmetry.
We have the freedom to choose the origin and direction of the Cartesian coordinate system in (\ref{eq:1-12a}) globally.
First, by considering the translational symmetry $x \to x+ x_0$ and $y \to y+y_0$, he concluded that $\delta$ could be left out.
Secondly, he considered $\SO(2):=\displaystyle{\left\{{\small{\begin{pmatrix} \cos \theta &-\sin \theta\\ 
\sin \theta& \cos\theta \end{pmatrix} }} | \theta \in \RR\right\}}$ 
$=\displaystyle{\left\{
\frac{1}{\sqrt{a^2+b^2}}{\small{\begin{pmatrix} a &-b\\ 
b& a \end{pmatrix} }}|a, b\in \RR\right\}}$ symmetry behind (\ref{eq:1-12a}) i.e., ${\small{\begin{pmatrix} \cos \theta &-\sin \theta\\ 
\sin \theta& \cos\theta \end{pmatrix}\begin{pmatrix} x\\ y \end{pmatrix}}}$.
He wrote the group action, $X=\displaystyle{\frac{\beta x- \gamma y}{\sqrt{\beta^2+ \gamma^2}}}$, 
$Y=\displaystyle{\frac{\beta y+\gamma x}{\sqrt{\beta^2+ \gamma^2}}}$.
Euler noticed that by replacing $X$ with $x$, we could set $\gamma = 0$.
}

\revs{
Here we show precisely why we can formally set $\gamma=0$.
By its inverse transformation, we have
\begin{equation}
\begin{pmatrix} x \\ y \end{pmatrix}
= \frac{1}{\sqrt{\beta^2+ \gamma^2}}
\begin{pmatrix} \beta & \gamma \\ -\gamma & \beta \end{pmatrix}
\begin{pmatrix} X \\ Y \end{pmatrix}, \ \ 
\begin{pmatrix} dx \\ dy \end{pmatrix}
= \frac{1}{\sqrt{\beta^2+ \gamma^2}}
\begin{pmatrix} \beta & \gamma \\ -\gamma & \beta \end{pmatrix}
\begin{pmatrix} dX \\ dY \end{pmatrix},
\end{equation}
Due to the property of $\SO(2)$, we have $d x^2 + d y^2 = d X^2 + d Y^2$.
Hence, the numerator of the left hand side in (\ref{eq:1-12a}) is equal
\begin{eqnarray*}
&&2\sqrt{\alpha\frac{\sqrt{dx^2+dy^2}+\beta dy + \gamma dx}{d x}}\\
&&=
2\sqrt{\alpha \frac{
\sqrt{\beta^2+ \gamma^2}\sqrt{dX^2 + dY^2}+\beta(-\gamma dX +\beta dY)+\gamma
(\beta dX + \gamma dY)}{\beta dX + \gamma dY}   }\\
&&=
2\sqrt[4]{\beta^2+ \gamma^2}\sqrt{\alpha \frac{
\sqrt{dX^2 + dY^2}+\sqrt{\beta^2+ \gamma^2} dY}{dX}}
\sqrt{\frac{dX}{\beta dX + \gamma dY}},
\end{eqnarray*}
while the denominator is given by
$$
\sqrt[4]{1+p^2}=
\sqrt{\frac{\sqrt{dX^2 + dY^2}}{dX}}
\sqrt{\frac{dX}{\beta dX + \gamma dY}}.
$$
By the rotation, (\ref{eq:1-12a}) is reduced to
$$
\frac{2\sqrt{\alpha\sqrt{(1+(dY/dX)^2)}+\sqrt{\beta^2+\gamma^2}(dY/dX)}}
{(1+(dY/dX)^2)^{1/4}}=\sqrt{\beta^2+\gamma^2} X.
$$
This guaranteed Euler's argument; we replace $X$ with $x$ and $\sqrt{\beta^2+\gamma^2}$ with $\beta$.
}

By showing $\delta=0$ and $\gamma=0$, he had the relation \cite[l.-9, p.249]{E2}\revs{\cite[l.-8, p.81]{OEB}},
\begin{equation}
2\sqrt{(\alpha\sqrt{(1+pp)}+\beta p}
=\beta x(1+pp)^{1:4}
\label{eq:1-13}
\end{equation}
and \cite[l.-8, p.249]{E2} \revs{\cite[l.-7, p.81]{OEB}},
\begin{equation}
\revs{4}\alpha\sqrt{(1+pp)}+\revs{4}\beta p=\beta^2 x^2\sqrt{(1+pp)}.
\label{eq:r-27}
\end{equation}
Noting 
$
\revs{4}\beta p=(\beta^2 x^2 -4\alpha)\sqrt{(1+p^2)},
$
he implicitly found
\begin{equation}
(\revs{(4\beta)^2}-(\beta^2 x^2 -4\alpha)^2)p^2 =(\beta^2 x^2 -4\alpha)^2.
\label{eq:Erstl}
\end{equation}
Then he introduced some parameters $\alpha = \dfrac{4m}{a^2}$ and $\beta = \dfrac{4n}{a^2}$, and finally obtained the nonlinear differential equation \cite[l.-4, -5, p.249]{E2} \revs{\cite[l.-6, p.81]{OEB}},
\begin{equation}
\frac{dy}{dx}=p=\frac{(nn xx -m aa)}
{\sqrt{(n^2a^4-(\beta^2 xx -maa)^2)}}.
\label{eq:1-14a}
\end{equation}

However, in this paper, \revs{
instead of these parameters $m$, $n$, and $a$, and we will continue to use $\alpha$ and $\beta$ to rewrite his result (\ref{eq:1-14a}) based on (\ref{eq:r-27}) and (\ref{eq:Erstl}) for later convenience.}
His final result (\ref{eq:1-14a}) is essentially the same as
\begin{equation}
\frac{dy}{dx}=p=\frac{(\beta^2 x^2 -4\alpha)}
{\sqrt{(\revs{(4\beta)^2}-(\beta^2 x^2 -4\alpha)^2)}}.
\label{eq:1-14}
\end{equation}
Further, recalling $ds = \sqrt{(1+p^2)} dx$, he essentially found
\begin{equation}
ds=\frac{\revs{4}\beta dx}
{\sqrt{(\revs{(4\beta)^2}-(\beta^2 x^2 -4\alpha)^2)}},
\label{eq:1-15}
\end{equation}
\begin{equation}
\left(
s=\int_{X_0}^X\frac{\revs{4}\beta dx}
{\sqrt{(\revs{(4\beta)}^2-(\beta^2 x^2 -4\alpha)^2)}}, \quad
y=\int_{X_0}^X\frac{(\beta^2 x^2 -4\alpha)}
{\sqrt{(\revs{(4\beta)}^2-(\beta^2 x^2 -4\alpha)^2)dx}}.\right)
\label{eq:1-16Ex}
\end{equation}

Then by shifting the origin of $x$ and renaming the parameters, Euler obtained
\begin{equation}
s=\int_{X_0}^X\frac{a^2 dx}
{\sqrt{(a^2-(\alpha +\beta x+\gamma x^2)^2)}}, \quad
y=\int_{X_0}^X\frac{(\alpha +\beta x+\gamma x^2)dx}
{\sqrt{(a^2-(\alpha +\beta x+\gamma x^2)^2)}},
\label{eq:1-16E}
\end{equation}
which are well-known results.

\revs{
Euler numerically evaluated the integrals (\ref{eq:1-16E}) as in Figure \ref{fg:1.2}, and he completely classified the elasticae; he implicitly determined the moduli of elliptic curves \cite{M3}.
}
\begin{figure}[htb]
\begin{center}
\includegraphics[width=0.8\hsize]{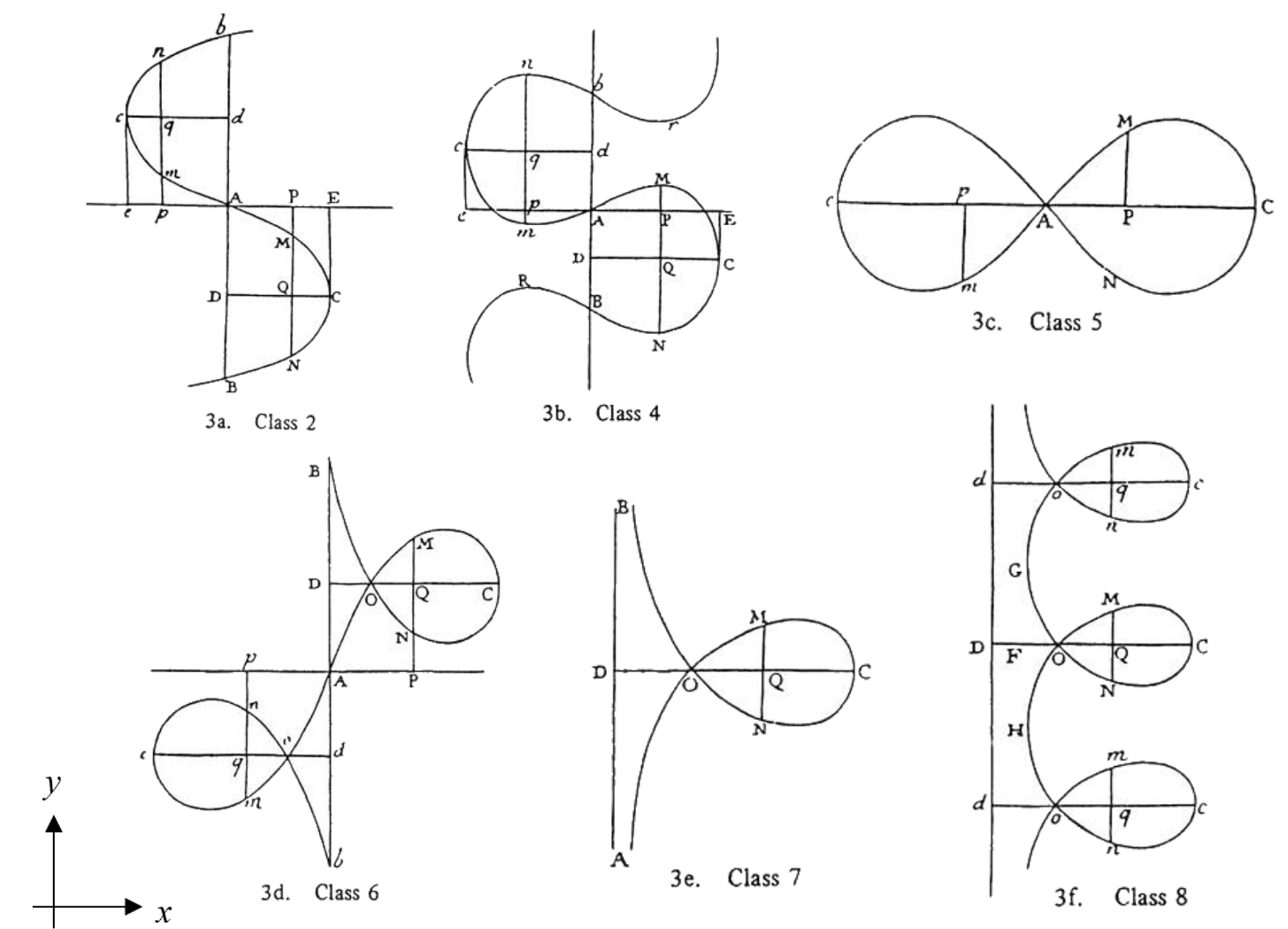}\\
\caption{Euler's diagrams of elastica:
Euler illustrated the shapes of the elasticae by numerical computations \cite{E2}}
\label{fg:1.2}
\end{center}
\end{figure}

\section{Discussion}

We showed the derivation of Euler's elastica equation.
In the derivation, we showed that he used Noether's theorem, which was published in 1918, and showed the governing equation, which differs from (\ref{eq:SGe}) and looks a complicated formula transformation.
Thus mathematical meaning of his derivation is quite difficult for the successors.
In fact, most of the books and papers \cite{Di, DA, Le, Love, T1, T2}, there \revs{was} no essential description of Euler's derivation.
There (\ref{eq:SGe}) \revs{has been} regarded as the governing equation of elastica, which Lagrange and Gauss showed (see Appendix \revs{A.1}).

\revs{Although traditionally there might seem to have existed the view that Euler's derivation of the equation of elastica was accidental and technical, and the view (\ref{eq:SGe}) and its derivation presented by Gauss had more essential than Euler's derivation or Euler's results, we believe that such views are not an appropriate (if indeed such views existed).
Carefully analyzing his derivation, we conclude that Euler had his profound insight into the elastica and derived his equation.}
We should clarify what Euler did for what purpose.

\revs{We make some additional remarks on Euler's derivation.}

\begin{enumerate}

\item \revs{
It should be noted that Euler minimized the Euler-Bernoulli energy $\displaystyle{\int \frac{1}{R^2} ds}$ under the isometric (isoperimetric) condition and the boundary condition, although the isometric condition has sometimes been disparaged: In Gauss' derivation, the condition was not treated explicitly as in Appendix A.1, since the isometric condition is suppressed by considering the variation of the tangential angle $\varphi$ as in Appendix A.2.
}

\revs{
In order to evaluate the minimum under the constraint, he used the so-called Lagrange multiplier; 
it was thought that the multiplier was introduced by Lagrange \cite{Jah}.
}

\item \revs{
In determining $\gamma=\delta=0$ in (\ref{eq:1-12a}), Euler explicitly used the group action of the euclidean move.
It should be noted that he explicitly recognized and understood the group action on this system and the symmetry.
Since the symmetry and the group action (including gauge symmetry) are noticed by Weyl \cite{W1, W2} and Klein \cite{Klein} from a modern point of view, Euler's insight was so earlier.
It is known that Noether studied her theorem at the request of Klein who considered that the group action governed geometry.
}

\item By (\ref{eq:1-10}) and (\ref{eq:1-13}), Euler implicitly found
the relation between the curvature $\kappa$ and $x$,
\begin{equation}
\kappa = \beta \frac{x}{2},
\label{eq:x=k}
\end{equation}
that shows the symmetry of the elastica as mentioned in \cite{M1}.

\item We also note that the elastica obeys the SMKdV equation \revs{as in Appendix A.2} \cite{M1}.
(\ref{eq:1-15}) essentially shows the SMKdV equation,
\begin{equation}
\revs{8}\beta \frac{d^2 x}{d s^2}\revs{+}\beta^3x^3 -\revs{4}\alpha\beta x  =0,
\label{eq:1-16a}
\end{equation}
\begin{equation}
\revs{8}\beta^2 \frac{d^3 x}{d s^3}+\revs{3}\beta^4x^2\frac{dx}{ds} -\revs{4}\alpha\beta^2 \frac{dx}{ds}=0.
\label{eq:1-16}
\end{equation}
If we put (\ref{eq:x=k}) into (\ref{eq:1-16a}), we \revs{recover} the SMKdV equation (\ref{eq:SMKdVeq0}),
\begin{equation}
a \kappa +\frac{1}{2} \kappa^3 + \frac{d^2}{d s^2}\kappa=0.
\label{eq:xSMKdV}
\end{equation}

\item As mentioned above, (\ref{eq:1-6}) should be regarded as the Noether current (\ref{eq:1-8}) \cite{Go, N}.
\revs{
(The {\lq\lq}current{\rq\rq} is the terminology in the elementary particle physics since the generalization of (\ref{eq:1-8}) means the conservation law of the  current including the electric current, e.g., $\frac{\partial}{\partial t} \cI_0-\frac{\partial}{\partial s} \cI_1=0$ \cite{Go}.)
More precisely, Euler investigated $\frac{d}{ds}\cI =\frac{d x}{d s}\frac{d}{dx}\cI=0$ in (\ref{eq:1-8}).
}
Euler's variation theorem includes Noether's theorem only for translational symmetry, whose general version for general group action was discovered by Noether in 1918 \cite{N}.

Thus, probably, Lagrange could not understand \revs{the} essentials of this derivation, (\ref{eq:1-14}) and (\ref{eq:1-15}), and thus he showed another elastica equation (see Appendix \revs{A.1}).

\smallskip
\color{black}
Naturally, the question arises why the Noether current is required in Euler's derivation.
As its answer, the origin of the requirement of the Noether current is that Euler used $x$ as the parameter of the curve $\{z(s)=x(s)+\ii y(s)$ $ |\ s\in [0, L]\}$.
The curve $z$ is a function of $s$, but $y$ is a multivalued function of $x$ as Figures \ref{fg:1.1} and \ref{fg:1.2}. (In the framework of modern mathematics, it is very difficult to consider $y$ as a function of $x$, but $\{z\}$ is a covering of an interval $[x_0. x_1]=\{x\}$: $\varpi : \{z\} \to [x_0, x_1]$: {\lq\lq}covering{\rq\rq} is a terminology in modern mathematics.
There are (infinite) $y$'s for each point $x$ in $[x_0, x_1]$. i.e., $\{y(x)\ |\ x\in [x_0, x_1]\}$. ). 

The energy minimum of $\displaystyle{\int (\kappa(x))^2 d s(x)}$ can lead to the family of minimized curves $y(x)$'s, so we should pick up a connected curve (piece) in the family of curves $\{y(x)\ |\ x\in [x_0, x_1]\}$: mathematically speaking, we get it from the inverse image $\varpi^{-1}$ of $[x_0, x_1]$.
The connected curve $z(s)$ is characterized by the translational symmetry $z(s)$ to $z(s+s_0)$ for any $s_0\in \RR$ via $s(x+\delta x)=s(x)+\delta s$ for the infinitesimal $\delta x$.
Thus Euler had to use the translational symmetry to obtain (\ref{eq:1-16E}).
\color{black}

\item
Since Euler's fluid equation is also connected with Noether's theorem for translational symmetry as in \cite{A, EM, MKS}, we should re-evaluate Euler's variational method based on this observation.
When he derived his fluid equation, there is a possibility that he used the idea related to Noether's theorem.

\revs{
Further, we note that the translational symmetry is given by $\Phi_{g_{s'}} z(s)=z(s+s')$ for the unitary group $g_{s'}\in \UU(1)=\{\ee^{\ii s}\}$ in Introduction.
Noting the arguments after (\ref{eq:1-12a}), Euler implicitly understood such a group action in physical objects.
The above is also realized for a regular function $z(s)$ by 
$\displaystyle{\ee^{t \frac{d}{ds}} z(s) =
\left[ \sum_{n=0}^\infty \frac{1}{n!} t^n \frac{d^n}{ d s^n}\right] z(s) = z(s+t)}$ by the Taylor expansion.
Accordingly, it is related to the equation 
$\displaystyle{\left(\frac{\partial}{\partial t}-\frac{\partial}{\partial s}\right)}f=0$, since formally we have $\displaystyle{\frac{\partial}{\partial t}\ee^{t \frac{\partial}{\partial s}}=\frac{\partial}{\partial s}\ee^{t \frac{\partial}{\partial s}}}$, i.e., $\displaystyle{\ee^{t \frac{\partial}{\partial s}}f(s)=f(s+t)}$ and thus $\displaystyle{\left(\frac{\partial}{\partial t}-\frac{\partial}{\partial s}\right)}f(s+t)=0$.
(These can be reinterpreted the correspondence between Lie group $\{\ee^{t\frac{d}{dt}}\}$ and Lie algebra $\{t \frac{d}{ds}\}$ in modern mathematics, e.g., \cite{AM}.)
}

\revs{
These observations remind us of Euler's investigation on the wave equation \cite{Bottazzini}.
}

\item Euler focused on the translational symmetry as $M=N=0$ and derived the Noether current.

The symmetry is related to the symmetry that we can choose any point in the elastica as the origin of $s$.
As we showed in Appendix \revs{A.3}, under the isometric condition based on the result of Goldstein and Petrich \cite{GP}, the translational symmetry provides the modified KdV hierarchy \revs{(\ref{eq:Hier2})} including the SMKdV equation \cite{MP}.
When we consider the excited states of elastica, the isometric deformations with the translational symmetry play the crucial role as the nature of elastica; More recently, it is closely related to the shape of supercoiled DNA due to thermal effect \cite{M24}. 
 
\smallskip
Euler's treatment in his derivation is quite natural from such a viewpoint.
He derived his equation in the class of the isometric \revs{(isoperimetric)} deformations of the curves, which are described by the modified KdV equation (\ref{eq:MKdVeq}).
Although the static sine-Gordon equation (\ref{eq:SGe}) has been known as the elastica equation, it is not related to such symmetries and thus to the excited states.
Therefore, (\ref{eq:SMKdVeq0}) is more profound than (\ref{eq:SGe}).
It is very surprising that Euler captured the essence of elastica.

In other words, from a modern point of view, his approach is more natural than that of Lagrange and Gauss, even though the multi-functions $\kappa$, $y$ and $s$ on $x$ are not proper.

\item In the derivation, Euler chose that $\gamma=0$.
It corresponds to break the symmetry between $x$ and $y$.
Due to the choice, we have (\ref{eq:x=k}).
We note that since $\kappa$ is bounded or periodic with respect to the arclength $s$, $x$ shows the periodic \revs{$x(s+L)=x(s)$,} whereas $y$ shows the quasi-periodic \revs{$y(s+L)=y(s)+y_L$ for a constant $y_L$ as Figures \ref{fg:1.1} and \ref{fg:1.2}.}

\revs{It is known that }
$x+\ii y$ is written well by the Weierstrass $\zeta$ function\revs{, $x+\ii y=a\zeta(bs +d)+cs$ for certain constants $a$, $b$, $c$ and $d$ \cite{M1}: we note that $x$ and $\dfrac{d}{ds} z$ are also expressed by the (periodic) elliptic $\wp$ function \cite{M1}, e.g.,
\begin{equation}
 \dfrac{d}{ds} z(s) = -a\,b\, \wp(bs+d) +c, \quad \dfrac{d}{ds} z(s+L)=\dfrac{d}{ds} z(s).
\label{eq:wp}
\end{equation}
} 
\revs{
From the viewpoint of history of the elliptic functions, Euler's results should be re-evaluated since his results were connected with the sigma function and moduli space of the elliptic curves as mentioned in Chapter 1 in \cite{M3}.}

\revs{
Since the $\zeta$ function is expressed by $\displaystyle{\frac{d}{du} \log \sigma(u)}$, where $\sigma$ is the elliptic sigma function which has the quasi-periodic properties $\sigma(u + \omega) = \ee^{\rho_\omega u+ \rho_0} \sigma(u)$ for certain constants $\omega$, $\rho_\omega$ and $\rho_0$, Euler's results on elastica showed the essence of the elliptic sigma function; the elliptic sigma function is crucial in modern elliptic function theory \cite{M3}.
It should be noted that Gauss found the prototype of the sigma function in 1800 by considering the static sine-Gordon equation as in \cite[Chapter 1]{M3}.
}

\item As in \cite{Bi}, Euler found the equation (\ref{eq:1-16E}) itself in 1732 \cite{E1}.
Thus his derivation was tuned for the discovery by following the suggestion of Daniel Bernoulli \cite{DA}.
We also note that (\ref{eq:1-16E}) was \revs{a} natural generalization of Jacob Bernoulli's expression of the rectangular elastica \revs{\cite{JB1694a},
\begin{equation}
dy = \frac{xx dx}{\sqrt{a^4-x^4}}, \qquad
ds = \frac{aa dx}{\sqrt{a^4-x^4}}.
\label{eq:Rectelas}
\end{equation}
}
While his derivation may have been perceived as an artificial manipulation, Euler understood the nature of elastica \revs{as mentioned above.}

\item Bolza reviewed Euler's derivation of his elastica equation in \cite{B}.
Since he studied instability problems related to isoperimetric problems \cite{B2}, which play the crucial role in the modern elastica problem \cite{CD}, and cited Born's study of the instability of elastica in \cite{B}, he understood the symmetry of elastica well. 
\revs{
Further, he was the expert of the Weierstrass elliptic function theory \cite[Chapter 1]{M3}.
}
However, he also considered the static sine-Gordon equation as the equation of elastica when he reviewed its derivation in \cite{B}, and thus he could not notice the symmetry hidden in the original derivation.

\end{enumerate}

\color{black}

\setcounter{section}{0}
\renewcommand{\thesection}{\Alph{section}}
\section{Appendix: static sine-Gordon equation and SMKdV equation}

\subsection{Elastica equation by Lagrange and Gauss}

\begin{figure}[htb]
\begin{center}
\includegraphics[width=0.37\hsize]{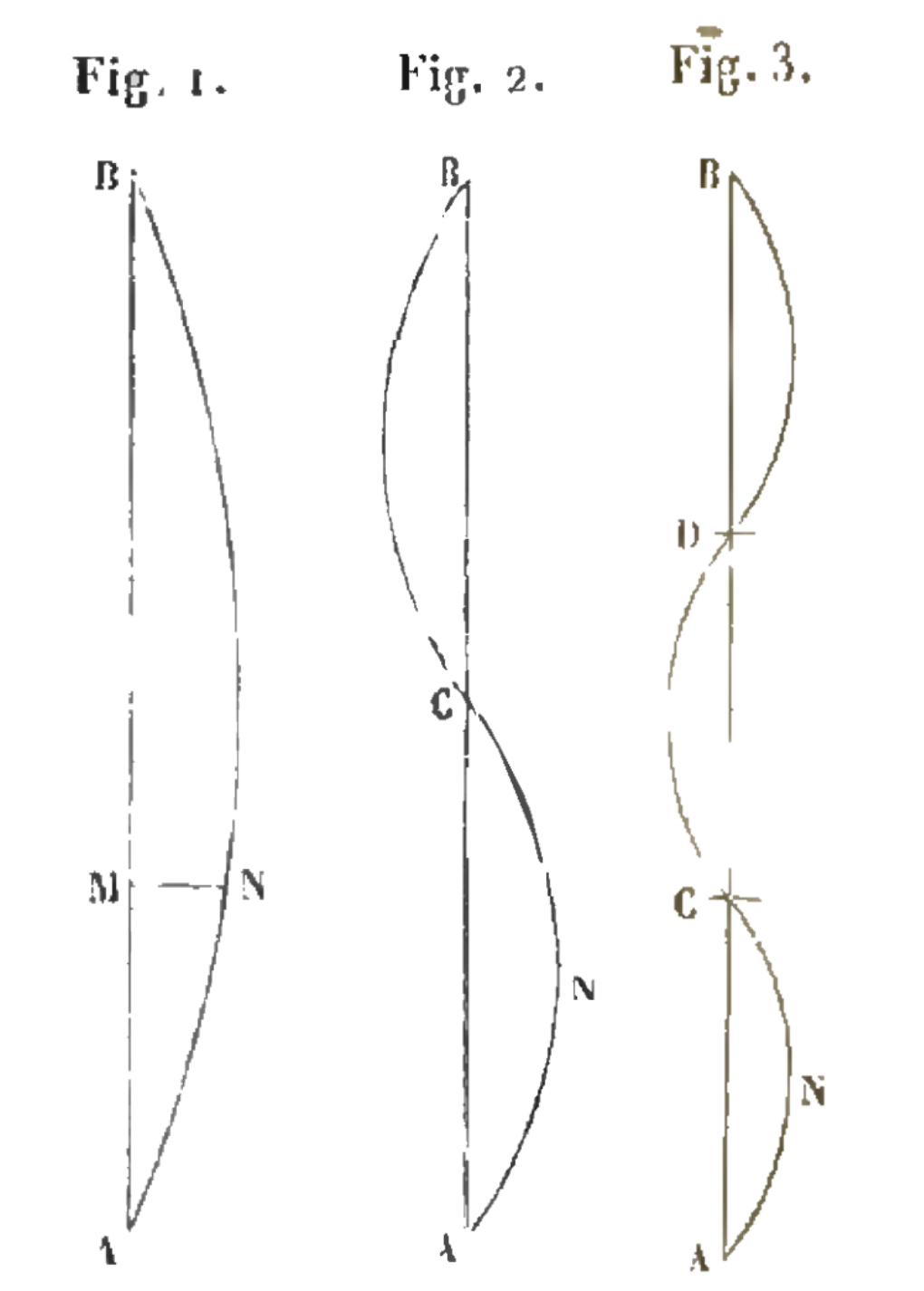}\\
\caption{Lagrange's diagrams of elastica \cite{La2}}
\label{fg:1.3}
\end{center}
\end{figure}

Euler found additional \revs{properties via the addition theorem} in the rectangular elastica integral \revs{(\ref{eq:Rectelas})}, or the lemuniscate integral\revs{; 
From the differential equation of the arclength $s$ and $s'$, $(m d s = n d s')$,
$$
\frac{m d x}{\sqrt{1-x^4}} = \frac{n d y}{\sqrt{1-y^4}},
$$
for positive integers $m$ and $n$, he found that $u$, $z$ and $w$ satisfying
$$
 w=\frac{z\sqrt{1-u^4}+u\sqrt{1-z^4}}{1+u u z z}
$$
implies that 
$$
\int^w_0 \frac{ d t}{\sqrt{1-t^4}} = \int^u_0 \frac{ d t}{\sqrt{1-t^4}}
+\int^z_0 \frac{ d t}{\sqrt{1-t^4}}.
$$
in 1752.
Subsequently,}
he studied the bending problem of elastica in 
{\it{Sur la force des colonnes}} in 1757;
Since in the bending problem, $1/2$ and $1/3$ play an essential role,  such an investigation might be natural after his discovery of the addition \revs{theorem} in the integral \revs{as in the diagrams of Lagrange in Figure \ref{fg:1.3}.}
Corresponding to it probably, Lagrange wrote a paper {\it{Sur la figure des colonnes}} in 1770 \cite{La2}, and wrote an equation of elastica \revs{ \cite[p.132]{La2},
$$
ds = \frac{d \varphi}{\sqrt{\frac{P}{L}- \frac{P^2}{4}\frac{f^2}{K^2}\cos^2\varphi}},
$$
 which is essentially the same as (\ref{eq:SGe}).}
In 1771, Lagrange investigated the relation between the arclength $s$ and the tangential angle $\varphi$ of the elastica, and he derived \cite[\S X, p.102]{La}
\begin{equation}
\frac{K^2d^2 \varphi}{d s^2} = C-P\cos\varphi+ N \sin \varphi .
\label{I.6.L10}
\end{equation}
By using it, he had the elliptic integrals .
\begin{eqnarray}
d s &=& \frac{K d \varphi}{\sqrt{\frac{M^2c^2}{4K^2}+P(1-\cos\varphi)+N\sin \varphi}}, \nonumber \\
d x &=& \frac{K \cos \varphi d \varphi}{\sqrt{\frac{M^2c^2}{4K^2}+P(1-\cos\varphi)+N\sin \varphi}}, \nonumber \\
d y &=& \frac{K \sin \varphi d \varphi}{\sqrt{\frac{M^2c^2}{4K^2}+P(1-\cos\varphi)+N\sin \varphi}}.
\end{eqnarray}

On the other hand, Gauss investigated the variational problem \cite{Ga}.
He noted $R = \displaystyle{\frac{ds}{d \varphi}}$ and considered
\begin{eqnarray}
&\delta& \int 
\left[\frac{1}{2}\left(\frac{d \varphi}{d s}\right)^2
-A \sin \varphi+B\cos\varphi\right]
 ds \nonumber \\
&=&\delta \varphi \frac{d \varphi}{d s}-
\int \delta \varphi 
\left[\frac{d^2 \varphi}{d s^2}
+A \cos \varphi+B\sin\varphi\right]
 ds = 0.
\label{I.6.10}
\end{eqnarray}
Then he found
\begin{equation}
\frac{dd \varphi}{ds^2}+A\cos \varphi+B\sin\varphi =0.
\label{I.6.11}
\end{equation}

These (\ref{I.6.L10}) and (\ref{I.6.11}) are well-known results (\ref{eq:SGe}) in the standard textbook of elastica \cite{Love}.
However, these are different from Euler's result, although due to the equivalent relations in the elliptic function theory, (\ref{I.6.L10}), (\ref{I.6.11}) and 
Euler's result (\ref{eq:1-16E}) are consistent.

\revs{
Here we remark that since the isometric (isoperimetric) condition is always satisfied in Gauss' derivation, as shown in the following note in the next section, it does not require attention.}

\subsection{The first derivation of the SMKdV equation}

It is also noted that the elastica obeys the SMKdV equation (\ref{eq:SMKdVeq0}).
We directly show it.
Let $z(s):= x(s)+ \ii y(s)$ and $s \in [0, L]$ for $L>0$, \revs{where $s$ is the arclength.
Let $\displaystyle{\frac{d}{ds}z(s) = \ee^{\ii \varphi}}$ be the tangent vector where $\varphi$ is the tangential angle.
We assume that $\varphi(s+L)=\varphi(s)$.
Thus, as long as we consider the infinitesimal variation $\varphi$ to $\varphi+\delta \varphi$, the isometric condition naturally holds, since $|\ee^{\ii (\varphi+\delta \varphi)}|=1$; it is not necessary to take care of the isometric condition explicitly in Gauss' derivation mentioned above.
}

The curvature $\kappa$ is given by $\kappa = \displaystyle{
\frac{1}{\ii} \frac{d}{d s} \log \frac{d z}{d s}}$.
We consider the variation of $z$, 
$$
z_\varepsilon (s_\varepsilon) = z(s) + \ii \varepsilon(s)\dfrac{d}{d s} z.
$$
For the isometric condition, we should estimate the contribution of the measure \revs{$ds$ in the variation.}
Since
$\displaystyle{
   \frac{d z_\varepsilon}{d s}  = (1 - \varepsilon \kappa (s)+
\ii\frac{d \varepsilon}{d s}  ) \frac{d}{d s} z
}$,
we have
$$
   d s_\varepsilon^2 = d \overline{z_\varepsilon}
d z_\varepsilon
      = (1 - 2 \varepsilon \kappa + O(\varepsilon^2)) d s^2.
$$

The curvature of $z_\varepsilon$, 
$\kappa_\varepsilon: =\displaystyle{-\ii \frac{d}{d s_\varepsilon}
\log \frac{d}{d s_\varepsilon} z_\varepsilon}$, is given by
$$
\kappa_\varepsilon = \kappa + 
\left[\kappa^2+\frac{d^2}{d s^2}\right] \varepsilon
      + O(\varepsilon^2).
$$
The Euler-Bernoulli energy functional is given by
$$
\kappa_\varepsilon^2 d s_\varepsilon
= \left[\kappa^2 + \left[\kappa^3+2\kappa\frac{d^2}{d s^2}\right] \varepsilon
      + O(\varepsilon^2)\right] d s.
$$
With the Lagrange multiplier $a$, we have the minimal of the Euler-Bernoulli energy functional,
\begin{equation}
    \frac{\delta (2\cE_\varepsilon-2a \int_{[0,L]} d s_\varepsilon)}
{\delta \varepsilon(s)}
       = \kappa^3 +2 \frac{d^2}{d s^2}\kappa + 2a \kappa = 0,
\end{equation}
or
\begin{equation}
    a \kappa +\frac{1}{2} \kappa^3 + \frac{d^2}{d s^2}\kappa=0,
\label{eq:SMKdV}
\end{equation}
which is the same as (\ref{eq:SMKdVeq0}) and (\ref{eq:xSMKdV}).

As mentioned in the third remark in Discussion, \revs{since} $\kappa$ is proportional to $x$ in the elastica, we reproduce Euler's result (\ref{eq:1-16a}).
\revs{
Here we also note that Euler's investigation contained the contribution of the deviation of the measure $ds$:
since the concept of the measure is, a little bit, difficult even in the modern mathematics and thus the care on the contribution seems to be difficult, (\ref{eq:SMKdV}) did not appear before the middle of the 20th century except Euler's result.
(We have encountered wrong results even now due to the lack of the notion on the measure.)
}

\subsection{The second derivation of the SMKdV equation}

We continue to use the same convention as the above subsection.
However, we consider a continuous deformation of a closed curve $z_t:S^1 \to \CC$ by $t \in [0, \epsilon)$, $(\epsilon>0)$, which is again simply denoted by $z$;
($t$ is physically connected with the Schwinger proper time in quantum field theory \cite{M24}.)
The deformation is given by real valued functions $u_1$ and $u_2$ on $S^1\times [0, \epsilon)$ by
$$
        \partial_t z = (u_1(s,t)+\ii u_2(s,t))\partial_s z,
$$
since $\partial_s z$ is the tangential vector of $z$ which generates the two dimensional vector space $\CC$ due to $|\partial_s z|=1$.
Here we use the convention $\partial_t = \partial/ \partial t$ and $\partial_s = \partial/ \partial s$.
Goldstein and Petrich introduced the non-stretching condition (isometric condition) $[\partial_t, \partial_s]=0$, and showed that the isometric deformation is given by the relations \cite{GP},
\begin{equation}
\partial_t \kappa = \Omega u_2, \quad u_1=\partial_s^{-1}\kappa u_2,\quad
\Omega:=\partial_s^2 + \partial_s \kappa \partial_s^{-1}\kappa ,
\label{eq:dtkOmega1}
\end{equation}
\begin{equation}
\partial_t \kappa = {\tOmega} u_1, \quad u_2=\frac{1}{\kappa} \partial_s u_1,\quad
{\tOmega}:=\partial_s^2\frac{1}{\kappa}\partial_s + \partial_s \kappa  ,
\label{eq:dtkOmega2}
\end{equation}
because
$\partial_t \partial_s z=\revs{\partial_t \ee^{\ii\varphi}=}\ii \partial_t \varphi \partial_s z$, while
$\partial_s \partial_t z=[(\partial_s u_1+\kappa u_2)+\ii(\partial_s u_2 - k u_1)] \partial_s z$.
Here $\partial_s^{-1}$ is the pseudo-differential operator.

We note that if you choose $u_1$, then $u_2$ is determined, and vice versa.
Thus we focus on $u_2$.

As in \cite[Prop. IV.3]{MP}, if two isometric deformations, $u_2$ and $u_2'$, i.e.,
$u_1=\partial_s^{-1}\kappa u_2$ and $u_1'=\partial_s^{-1}\kappa u_2'$ \revs{with}
\begin{gather*}
\begin{split}
\partial_{t} z  &= (u_1+\ii u_2)\partial_s z \quad 
\mbox{or}\quad
\partial_t \kappa = \Omega u_2,\\
\partial_{t'} z  &= (u_1'+\ii u_2')\partial_s z \quad
\mbox{or}\quad
\partial_{t'}\kappa = \Omega u_2',\\
\end{split}
\end{gather*}
satisfy the relation $\partial_t \kappa = u_2'$, then
the deformation $\partial_{t} z$ is isoenergy deformation, i.e.,
$ \partial_t \displaystyle{\int \frac{1}{2}\kappa^2 ds = 0}$, and 
$\partial_{t'} \kappa =\Omega \partial_t \kappa$. 

The proof is simple.
$\partial_t\kappa^2=\kappa \partial \kappa = \kappa u_2'=\partial_s u_1'$ due to the assumption and (\ref{eq:dtkOmega2}).
Then we find a hierarchy of the differential equations as in \cite{MP}, \revs{i.e.,
\begin{equation}
 \partial_{t_i} \kappa = \Omega^i u_2,\quad 
\mbox{for }i=0,1,2,\ldots,
\label{eq:Hier1}
\end{equation}
where $(t_i)_{i=0, 1, \ldots}$ is the infinite parameters.}

We consider \revs{the} $u_2=0$ case, i.e., $\partial_t z= \partial_s z$, and $\partial_t \kappa = \partial_s \kappa$, which means the translational symmetry that Euler essentially used.
It is a trivial isoenergy and isometric {\lq\lq}deformation.\rq\rq

In other words, since the translational action or changing the origin of $s$, $\partial_t z= \partial_s z$, cannot change the energy or local length, it is natural to consider the isometric deformation with respect to $t'$ that guarantees the trivial conservation, which is the modified KdV equation (the differential of (\ref{eq:MKdVeq0}) in $s$ again),
\begin{equation}
 \partial_{t'} \kappa = \Omega \partial_s \kappa = \frac{1}{2}\partial_s \kappa^3 +\partial_s^3 \kappa.
\label{eq:MKdVeq}
\end{equation}
(Since we have
$$
\partial_{t'} \int\frac{1}{2}\kappa^2 ds = 
\int\kappa \left[\frac{1}{2}\partial_s \kappa^3 + \partial_s^3 \kappa\right] ds
=\int\partial_s \left[\frac{1}{8}\partial_s \kappa^4 + \kappa(\partial_s^2 \kappa -\frac{1}{2}(\partial_s \kappa)^2\right] ds,
$$
which must vanish, the deformation with respect to $t'$ also does not change the Euler-Bernoulli energy.
Thus we have the MKdV hierarchy\revs{,
\begin{equation}
 \partial_{t_i} \kappa = \Omega^i u_2,\quad 
\mbox{for }i=0,1,2,\ldots,
\label{eq:Hier2}
\end{equation} 
by putting $u_2=0$ into (\ref{eq:Hier1}), and infinite axes $(t_i)_{i=0, 1, \ldots}$ } showing the excited states of elastica and the shape of supercoiled DNA \cite{M24}.
\revs{
These parameters can be regarded as the Schwinger proper times in the context of the quantum field theory, which represent the geometry in the excited states of the elastica \cite{M24}.)}

Here we are\revs{, now,} concerned with a static equation that is consistent with (\ref{eq:MKdVeq}).
The deformation should be only for the translational direction of $s$.
Thus we put $t'=s/a$ in (\ref{eq:MKdVeq}), or   $\partial_{t'} \kappa =a \partial_s \kappa$, and then we recover the SMKdV equation (\ref{eq:SMKdV}) and (\ref{eq:xSMKdV}) by integrating it once.

This is the elegant derivation, although it is not directly related to the minimal problem.
\revs{
In other words, the isometric (isoperimetric, non-stretching) condition and the translational symmetry are much more important than the Euler-Bernoulli energy since both generate the Euler-Bernoulli energy as mentioned above.
The reason why both generates it may be related to the fact that the Euler-Bernoulli energy is the simplest harmonic energy $\displaystyle{\int \left(g^{-1} \frac{d}{ds} g\right)^2 ds}$ of $g=\ee^{\ii \varphi(s)}$ as a section of the $\UU(1)$-principal bundle (as the $\UU(1)$ gauge field in physical language), which is a purely geometric object \cite{M3}.
}
However, it is important to associate Euler's derivation with this one.
Euler focused on the gauge symmetry of $x$ and $y$ or $M=N=0$, which corresponds to the translational symmetry above.

\bigskip

\subsection*{Acknowledgments:}
The author is grateful to the anonymous referee for critical and valuable comments on the previous manuscript.
The author thanks Francesco Dal Corso for valuable discussions of the elastica problem and showing \cite{B2}.
The author has been supported by the Grant-in-Aid for Scientific Research (C) of the Japan Society for the Promotion of Science Grant, No.21K03289.

%
%
%


\bigskip
\bigskip

\noindent
Shigeki Matsutani\\
Electrical Engineering and Computer Science,\\
Graduate School of Natural Science \& Technology, \\
Kanazawa~University,\\
Kakuma Kanazawa, 920-1192, Japan\\
\texttt{ORCID:0000-0002-2981-0096}\\
\texttt{s-matsutani@se.kanazawa-u.ac.jp}

\end{document}